# In which fields can ChatGPT detect journal article quality? An evaluation of REF2021 results

Mike Thelwall
Information School, University of Sheffield, UK. https://orcid.org/0000-0001-6065-205X
m.a.thelwall@sheffield.ac.uk
Abdallah Yaghi
Information School, University of Sheffield, UK. https://orcid.org/0000-0002-9056-5579

**Abstract**
Time spent by academics on research quality assessment might be reduced if automated approaches can help. Whilst citation-based indicators have been extensively developed and evaluated for this, they have substantial limitations and Large Language Models (LLMs) like ChatGPT provide an alternative approach. This article assesses whether ChatGPT 4o-mini can be used to estimate the quality of journal articles across academia. It samples up to 200 articles from all 34 Units of Assessment (UoAs) in the UK's Research Excellence Framework (REF) 2021, comparing ChatGPT scores with departmental average scores. There was an almost universally positive Spearman correlation between ChatGPT scores and departmental averages, varying between 0.08 (Philosophy) and 0.78 (Psychology, Psychiatry and Neuroscience), except for Clinical Medicine (rho=-0.12). Although other explanations are possible, especially because REF score profiles are public, the results suggest that LLMs can provide reasonable research quality estimates in most areas of science, and particularly the physical and health sciences and engineering, even before citation data is available. Nevertheless, ChatGPT assessments seem to be more positive for most health and physical sciences than for other fields, a concern for multidisciplinary assessments, and the ChatGPT scores are only based on titles and abstracts, so cannot be research evaluations.

## Introduction

Evaluating the quality of other researchers' outputs is an important task for those involved in academic appointment, promotion, and tenure decisions. In many countries, including Australia, Italy, the UK, and New Zealand, there are also nation-wide period formal exercises to assess the quality of academic outputs to direct block grant funding (e.g., Buckle & Creedy, 2024; Franceschini & Maisano, 2017; Hicks, 2012; Sivertsen, 2017). Other countries sometimes evaluate research units in different ways, such by assessing individual research-intensive units to make a budget renewal decision or by investigating all departments in a discipline periodically (Iping et al., 2022), or by evaluating research only as part of broader institutional evaluations (Geuna & Martin, 2003). This consumes an enormous amount of expert time (Aczel et al., 2021) because evaluating the research quality of complex and unique outputs is difficult. This has led, in part, to the emergence of the field of scientometrics, with a focus on quantitative research evaluation, and many attempts to assess whether and when citation-based indicators could inform or replace human judgement. The consensus now seems to be that these indicators can inform human judgment in the health and physical sciences and to a weak extent in the social sciences and engineering, but not in the arts and humanities (e.g., Thelwall et al., 2023a). For this to be useful and relatively fair, at least three years of citation data may be needed (Wang, 2013), which is another substantial limitation in practice.

The emergence of Large Language Models (LLMs) suggests an alternative to citation analysis in the form of Artificial Intelligence (AI) for indicating or estimating the quality of academic research. Although machine learning has already been tried in scientometrics, it has either been used to predict long term citation counts (Qiu & Han, 2024) or has used citation data as inputs (Thelwall et al., 2023b) to estimate research quality, giving it similar limitations to citation-based indicators for research quality (Wilsdon et al., 2015). Many studies have shown that ChatGPT can provide useful support in the pre-publication peer review process by often making similar comments to reviewers or suggestions that reviewers would find helpful (Biswas et al., 2023; Liang et al., 2024b; Tyser et al., 2024), or by creating meta-reviews (Santu et al., 2024), although it also brings ethical and integrity challenges (Kim, 2024). Moreover, ChatGPT seems to be reasonably accurate at extracting specific information from academic publications, such as that needed for systematic literature reviews (Tao et al., 2024). Thus, it is logical to assess whether ChatGPT can be useful in research quality assessment.

Recent small-scale studies have now shown that ChatGPT can statistically significantly estimate the quality of a small set of variable quality journal articles from a single author (Thelwall, 2024ab), the quality ranking of 11 accepted and 10 rejected submissions to a single journal (ChatGPT 35, but not 4o: Saad et al., 2024), predict long term citation counts (de Winter, 2024), predict the conference committee decision for papers submitted to a computer science conference (Zhou et al., 2024), and be useful to identify potentially weak grant submissions for a funding agency (Carbonell Cortés, 2024). Nevertheless, there has been no attempt to assess whether ChatGPT quality predictions are reasonable for any academic field, so it is not clear whether the existing results generalise beyond narrow contexts. This article fills this gap with a single main research question and a secondary question for those considering using ChatGPT across multiple fields.
- RQ1: Can ChatGPT estimate the quality scores of journal articles in any or all academic fields?
- RQ2: Does ChatGPT have a field bias in the sense of giving higher quality scores to articles from some fields than others, relative to human reviewers?

## Methods

The research design was, for each of the UK's 34 broadly field-based Units of Assessment (UoAs – see figures below for names), take a sample of 200 journal articles submitted to REF2021 from the highest and lowest scoring sets of submissions to compare ChatGPT predictions with departmental average scores. Each of these sets is approximately a Higher Education Institution (HEI) department, centre, or institute, so the term "department" will be used for simplicity, even though "submission" is the usual REF term. The assessment was performed with correlation to test the extent to which the ChatGPT quality score alighed with the departmental average. It is not possible to correlate directly with article REF2021 scores because these are not published. Departmental average scores are used here as a substitute. This is appropriate because a department with a higher average score will tend to have higher scores for its individual articles. 100 articles were selected from the department(s) with the highest and 100 from the department(s) with the lowest average score to maximise the chance that the submission average score would reflect well the individual submission scores. Random numbers were used for article selection when there were more than 100 qualifying articles.

*Data*

The UK REF2021 assessed 185,594 outputs, most of which were journal articles. As part of a previous project (Thelwall et al., 2023b), these were matched with Scopus records by DOI (most) and (occasionally) by title, journal, and manual checking. Articles not in Scopus were discarded. Scopus was used as the source of the abstracts for the articles. These abstracts were cleaned to remove copyright statements and standard headings in structured abstracts. The latter was probably not necessary but might provide more concise and natural input information and reduces the chance that ChatGPT's results are due to leveraging journal style information.

The average institutional scores for departments are published on the REF2021 website but the individual article scores were destroyed as a matter of policy before the results were published. We could have used the average scores on the REF2021 website but as part of a previous project we had access to individual journal article scores and had calculated departmental average scores for those matching Scopus, so used these instead as they match the data used here exactly. The average departmental scores for articles in the top set of articles (usually 100) varied between 2.9 and 3.9, with an average of 3.5. The average departmental scores for articles in the bottom set of articles varied between 2.0 and 2.9, with an average of 2.5. The average departmental score difference between the top and bottom set of articles varied between 0.27 and 1.67, with an average of 0.99.

Within each UoA, the journal articles were ranked first by departmental average score (as above) and then by a random number and the top and bottom 100 were taken, with their title, abstract, and institutional average score. When there were less than 200 journal articles, all were included (UoA 28: 198; UoA 29: 93; UoA 31: 157). The articles and titles were all manually checked for errors and cleaned. These errors included incorrect hyphenation, apparently picked up by Scopus from line end hyphenation in abstracts, and incorrect copyright statement elimination. The titles and abstracts were converted to JSONL format for submission to the ChatGPT API, after shuffling them into a random order with a random number generator.

Article full texts were not sought for three reasons. First, previous research suggests that ChatGPT gives more useful research quality scores when fed with article titles and abstracts than if fed with titles alone or with articles, title, and full texts (Thelwall, 2024b). Second, not all full texts were available. Third, processing full texts from a variety of sources (mainly PDF) is error prone and a substantial labour-intensive task, even with automated assistance (e.g., the Python PyMuPDF package) (Bui et al., 2016; Stricker & Scheurer, 2023). Whilst it would have been useful to check that full texts would not improve the results, it was therefore impractical both for this project and most applications.

*ChatGPT setup*

The ChatGPT API was used instead of the ChatGPT web interface because of copyright restrictions. Whilst the web interface learns from its inputs, the API interface does not and so uploading articles and abstracts to the API does not indirectly violate article copyright. UK law also allows processing copyright material with machine learning for research, if the material has been lawfully accessed (https://www.legislation.gov.uk/ukpga/1988/48/section/29A, see also: Hawkes, 2012).

The task of research evaluation is complex and there are no agreed criteria, protocols or scoring systems. Nevertheless, rigour, significance, and originality are usually thought to be important or the main factors (Langfeldt et al., 2020), including in the REF. The UK REF2021 guidelines were therefore adopted for ChatGPT. These have the additional advantage that they closely match the instructions that the REF2021 evaluators had been told to follow. The quality scoring system is as follows (REF, 2019):
- 4*: "Quality that is world-leading in terms of originality, significance and rigour."
- 3*: "Quality that is internationally excellent in terms of originality, significance and rigour but which falls short of the highest standards of excellence."
- 2*: "Quality that is recognised internationally in terms of originality, significance and rigour."
- 1*: "Quality that is recognised nationally in terms of originality, significance and rigour."

There are four different guidelines for human reviewers scoring research outputs, one for each of the four main panels (A, B, C, D) that the 34 UoAs are grouped into. Each of these four guidelines was converted into a ChatGPT system input which describes the task (see Appendix). The original wording was reformulated to mimic the language used in the ChatGPT examples because this strategy seemed most likely to work well in the system. Each ChatGPT API session then consisted of submitting the system prompt, then a user prompt starting with "Score this journal article:", followed by the article title, the word "Abstract", and the abstract, separated by newline delimiters. ChatGPT 4o-mini was used with the default parameters because alternative parameters do not seem to improve the results (Thelwall, 2024b).

From previous studies, ChatGPT gives better results if it is queried many times and the average of its results is used (Thelwall, 2024ab). Thus, each set of up to 200 articles was submitted 30 times to ChatGPT consecutively (i.e., articles 1 to 200, then the same again 29 times).

The ChatGPT output is a set of paragraphs that almost always include a statement of the REF score. The exceptions are outputs with statements of the originality, rigour, and significance scores but not an overall score. A set of information extraction rules was constructed to identify the scores in these outputs, returning the average of the three separate scores, when an overall score was not given. The score was almost always a whole number but when it was a fraction (or when averaging the three independent scores produced a fraction), then this was used instead. When the rules could not find a score, the system prompted the first author to identify the score from the output text (see the AI menu of: https://github.com/MikeThelwall/Webometric_Analyst).

*Analysis*

The ChatGPT scores for each article, averaged over all 30 iterations, were correlated with the departmental average score for the submitting department to assess the extent to which ChatGPT was able to estimate research quality. Spearman correlations were used because the data is naturally ranks, even though averages and partial rank positions were included. Confidence intervals were calculated from the data produced with the same process as previously (Thelwall, 2024b) (the t-distribution formula or bootstrap data sampling).

For comparison with the main results, Spearman correlations were also calculated between departmental average scores and individual article scores for each UoA to compare with the ChatGPT correlations. These can be calculated from the data on the REF2021 website because it reports the number of outputs with each score (but not which output has any given score). For this we instead used equivalent data calculated from just the journal articles matched in Scopus, calculated from a previous project, for a more exact match. This is a more suitable calculation for many social sciences, arts and humanities because it excludes monographs, book chapters and other outputs that could tend to be their best outputs.

A bootstrapping approach was needed to estimate the Spearman correlation between (unknown) individual article REF scores and (known) departmental average REF scores. Although individual article scores are not available, the numbers of outputs of all types (e.g., including books) at each quality level are published on the REF2021 website (a spreadsheet in: https://results2021.ref.ac.uk/) and were used to approximate this. These data should work well for lower numbered UoAs, where journal articles dominate, but may work less well for some social sciences, the arts and humanities. The correlation between article REF scores and departmental ref scores for the samples for each UoA (usually 200 articles) was estimated by bootstrapping: randomly sampling (without replacement) articles from the REF-reported score distributions from each department the number of articles from that department submitted to ChatGPT. To give a simple example, if UoA 1's ChatGPT sample consisted of 100 articles from department A, which had 500 4* outputs and 500 3* outputs and a departmental average of 3.5*, and 100 from department B, which had 200 1* outputs and a departmental average of 1* then a random sample of 200 for correlation might be (4,3.5) x 49, (3,3.5) x 51, (1,1) x 100. This was repeated 1000 times and the average taken. In theory, if ChatGPT's scores were 100% correct for any UoA and ChatGPT's predictions were independent of the quality of the submitting department (which it was not directly told), then the ChatGPT correlation with institutional REF average scores would exactly match this bootstrapped correlation. More realistically, the closer the ChatGPT correlation is to the bootstrapped correlation, the more ChatGPT scores tend to align with the (unknown) actual REF scores.

# Results

## *Averages with different numbers of iterations*

For Main Panel A (mainly health and life sciences), except for UoA 1, the correlation between ChatGPT average scores and departmental average REF scores increased as the number of iterations increased and was moderate or strong (Figure 1). The UoA 4 Psychology, Psychiatry and Neuroscience correlation of nearly 0.8 is extremely high but even the UoA 6 Agriculture, Food and Veterinary Science correlation is moderate. The UoA 1 exception was an anomaly for the entire study and was also the first UoA sent to ChatGPT so, to check for experimental error, it was repeated after regenerating the dataset (so a partly different set of 200 articles) with almost identical results (not shown).

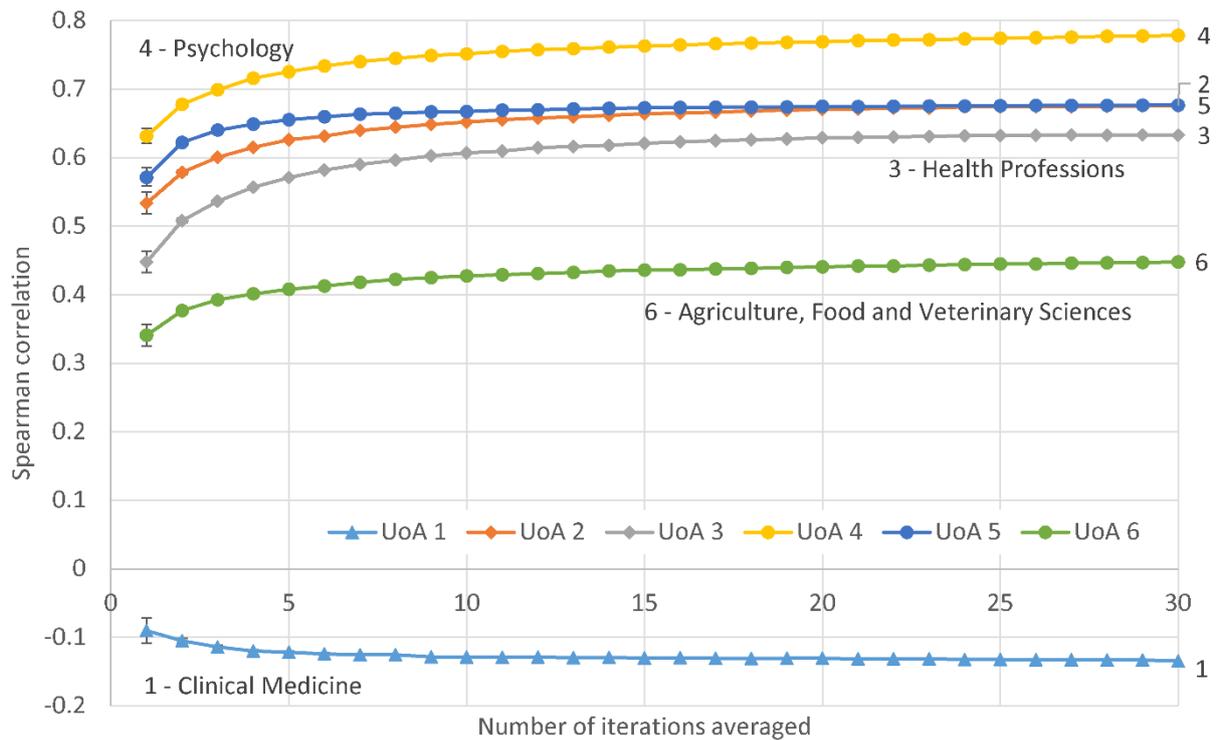

Figure 1. ChatGPT predictions averaged over n iterations correlated against departmental average REF scores for journal articles for Main Panel A UoAs (Medicine, Health and Life Sciences). Error bars indicate 95% confidence intervals.

For Main Panel B (physical sciences and engineering), the pattern was like that for Main Panel A above, but without an anomaly and with generally weaker correlations. The weakest correlation was for UoA 10 Mathematical Sciences, perhaps because its submissions were sometimes too complex or esoteric to be effectively processed by ChatGPT, or because it had different levels of confidence when assessing pure maths compared to applied maths and statistics within this single UoA.

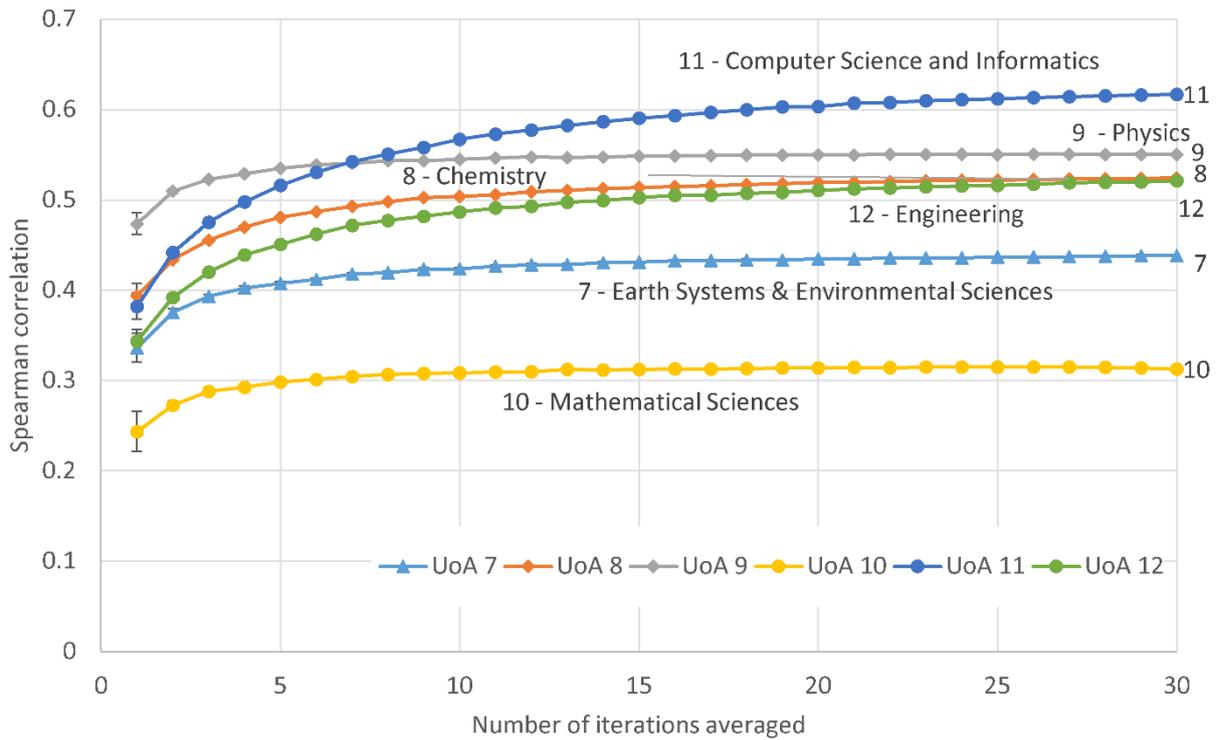

Figure 2. ChatGPT predictions averaged over n iterations correlated against departmental average REF scores for journal articles for Main Panel B UoAs (Physical Sciences, Engineering and Mathematics). Error bars indicate 95% confidence intervals.

For Main Panel C (social sciences), all correlations were again positive, and none were very weak (Figure 3). The lowest correlations seem to be for the more humanities-oriented fields. The slope of some of the lines suggests that higher correlations might have been obtained from larger numbers of iterations.

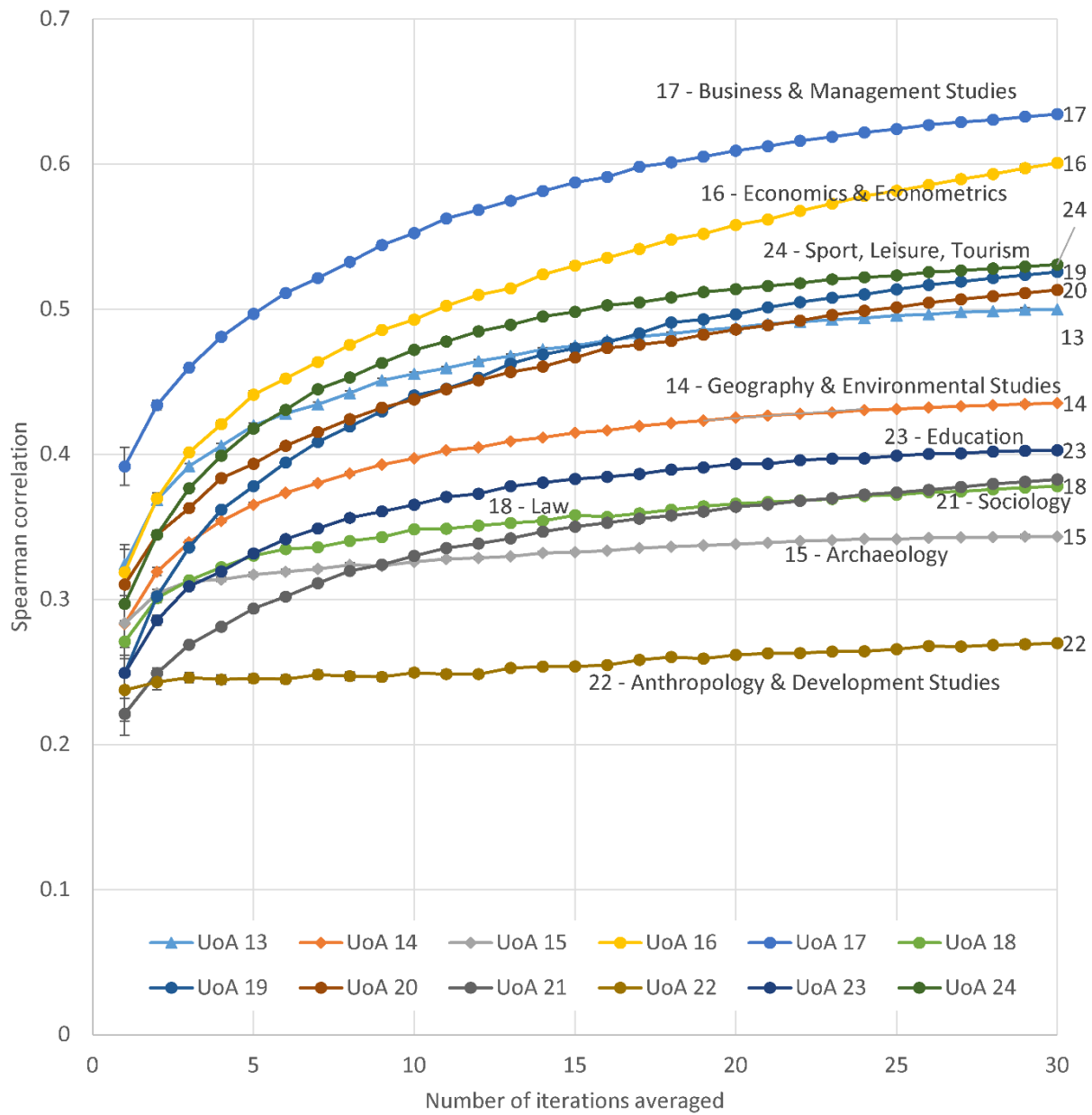

Figure 3. ChatGPT predictions averaged over n iterations correlated against departmental average REF scores for journal articles for Main Panel C UoAs (Social Sciences). Error bars indicate 95% confidence intervals.

For Main Panel D (arts and humanities), the correlations between ChatGPT scores and departmental average scores were the weakest overall, but still all positive. Within this, there is also a set of four very weak correlations, although no obvious common factor for them. It seems counterintuitive that some correlations decrease as the number of iterations increase, especially given the minute 95% confidence intervals, so the calculations were repeated in a different programming environment (R written by ChatGPT instead of VB.net written by the first author), to guard against programming bugs and identical results were obtained.

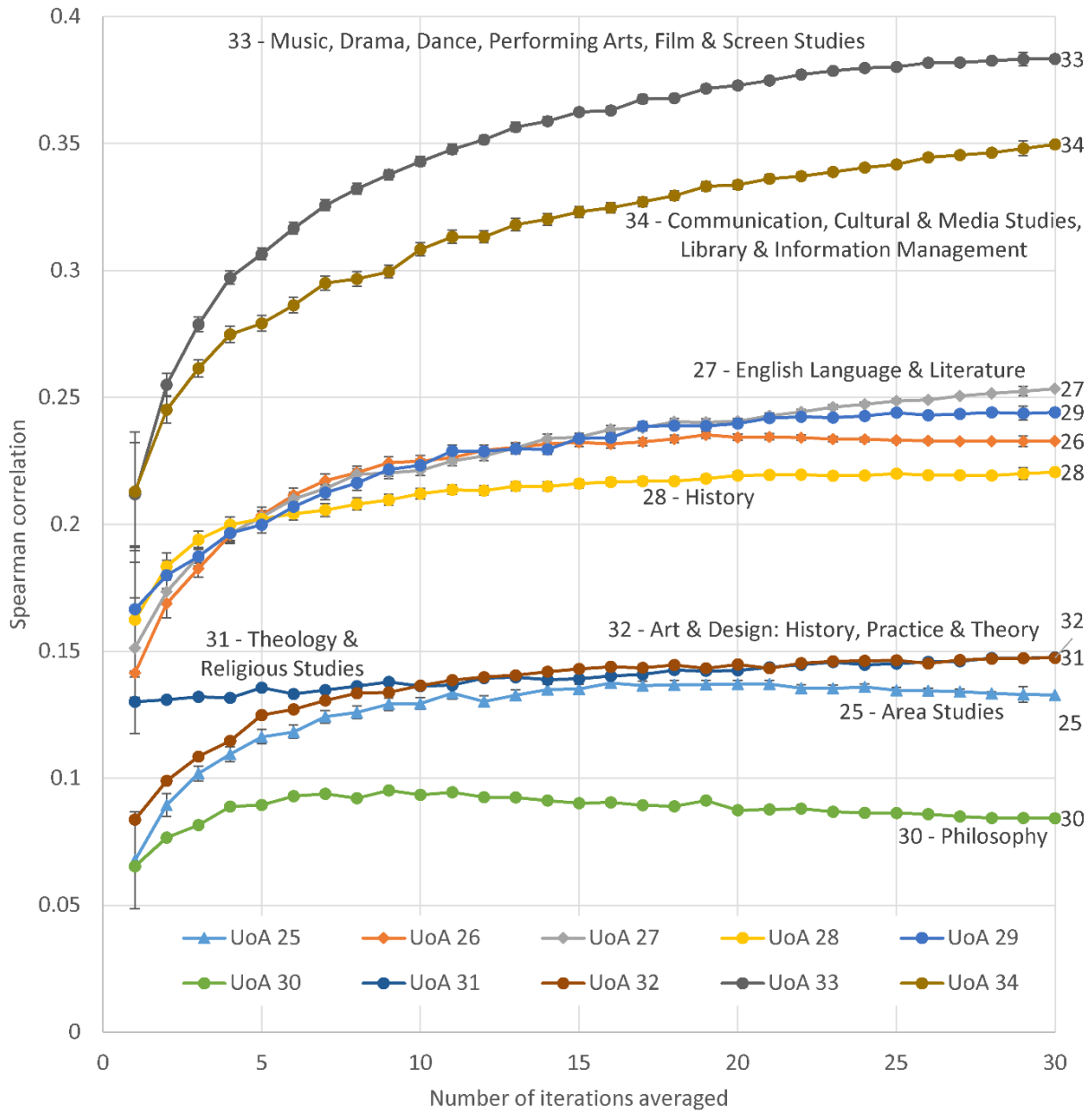

Figure 4. ChatGPT predictions averaged over n iterations correlated against departmental average REF scores for journal articles for Main Panel D UoAs (Arts and Humanities). Error bars indicate 95% confidence intervals.

## ChatGPT vs departmental average scores against article vs. departmental average scores

Recall that the correlations reported above are indirect in the sense of comparing the article-level score from ChatGPT with the departmental average REF score. The theoretical maximum correlation for ChatGPT should therefore be the correlation between the individual article scores and the departmental average scores. The latter was estimated through bootstrapping, as described in the methods (using all output types, not just journal articles).

As expected, the correlation between the ChatGPT scores and the departmental average scores is almost always lower than the estimated correlation between the individual REF scores and the departmental average scores (Figure 5). The three exceptions (UoAs 15, 27, 29) are within the confidence intervals of not being exceptions.

In many cases, and especially for the lower numbered UoAs, the ChatGPT correlation is very close to the theoretical maximum correlation for perfect predictions. Although the confidence intervals must be taken into account, this implausibly suggests that ChatGPT averages might very closely align with (rank the same as) article level scores in these UoAs. A possible alternative explanation is that higher scoring departments in some UoAs were more consistently able to make a case for the quality of their work in abstracts (e.g. by departmental REF training, policy or guidance for this issue) than to produce higher quality work. This is discussed in more detail below, together with other possibilities.

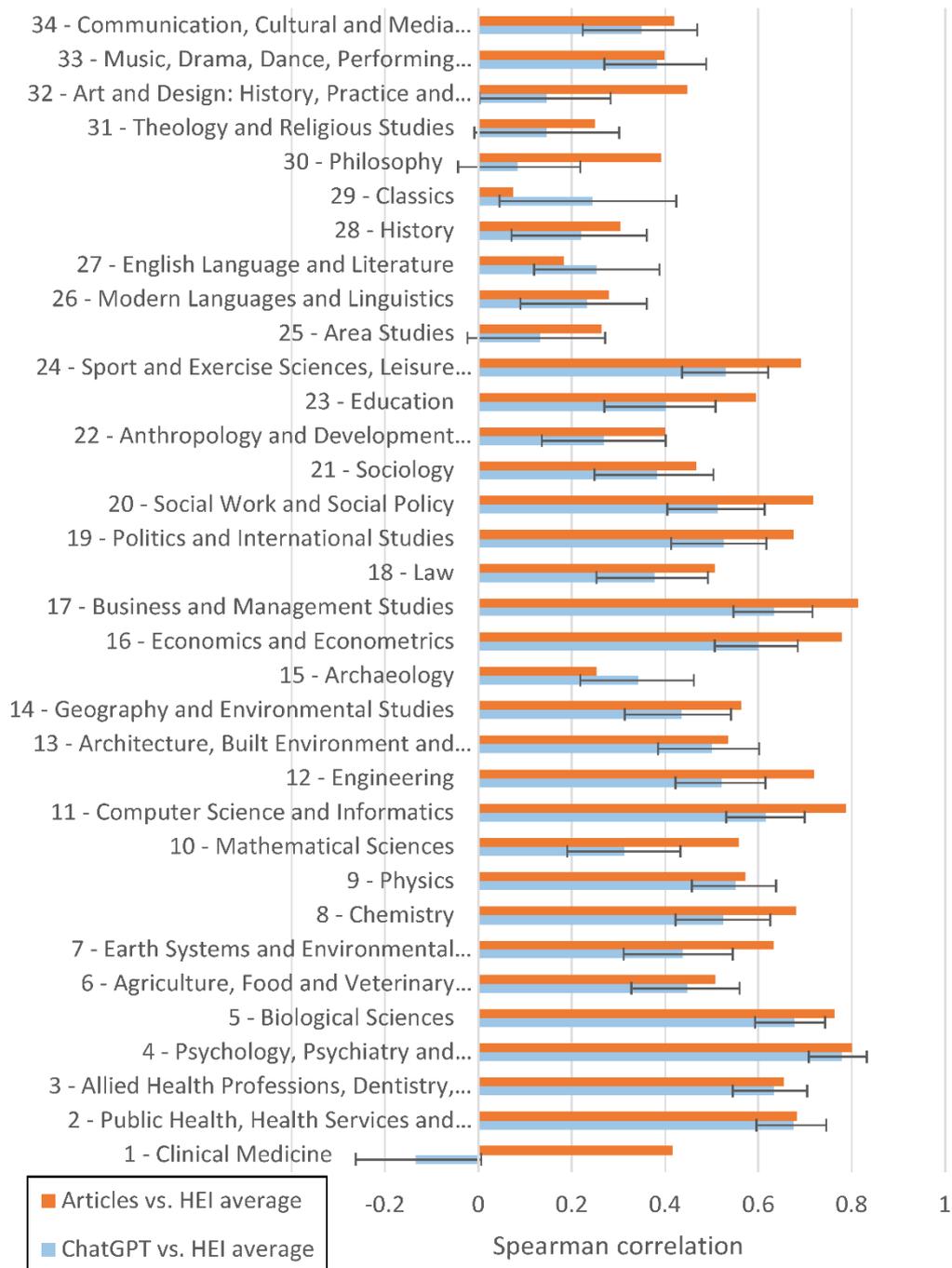

Figure 5. ChatGPT predictions averaged over 30 iterations correlated against departmental (HEI) average REF scores for journal articles for each UoA, showing 95% confidence intervals

for the population mean. Also included are estimates of the correlations between individual article scores and departmental average REF scores.

Population 95% confidence intervals for the correlation between the ChatGPT average scores and the departmental average scores were calculated with bootstrapping (Figure 5). These show the likely range of the theoretical overall correlation between departmental average scores and ChatGPT scores (for samples from high and low scoring departments), based on the specific samples of up to 200 analysed here. The overlaps between these show that in many cases the apparent greater power of ChatGPT for one UoA than another could be due to the samples selected, although it is less likely that panel-level trends are due to this.

### *ChatGPT average scores compared to human average scores*

ChatGPT noticeably tends to overestimate article scores, compared to the expert REF reviewers, in UoAs 1-9, the health and natural sciences, and a few other UoAs (e.g., Geography, Archaeology, History, Theology). In contrast, it underestimates them mainly in architecture and sport, with the other averages being similar (Figure 6). An underlying factor might be a ChatGPT preference for the definiteness of quantitative research, particularly when deciding if the top score is merited.

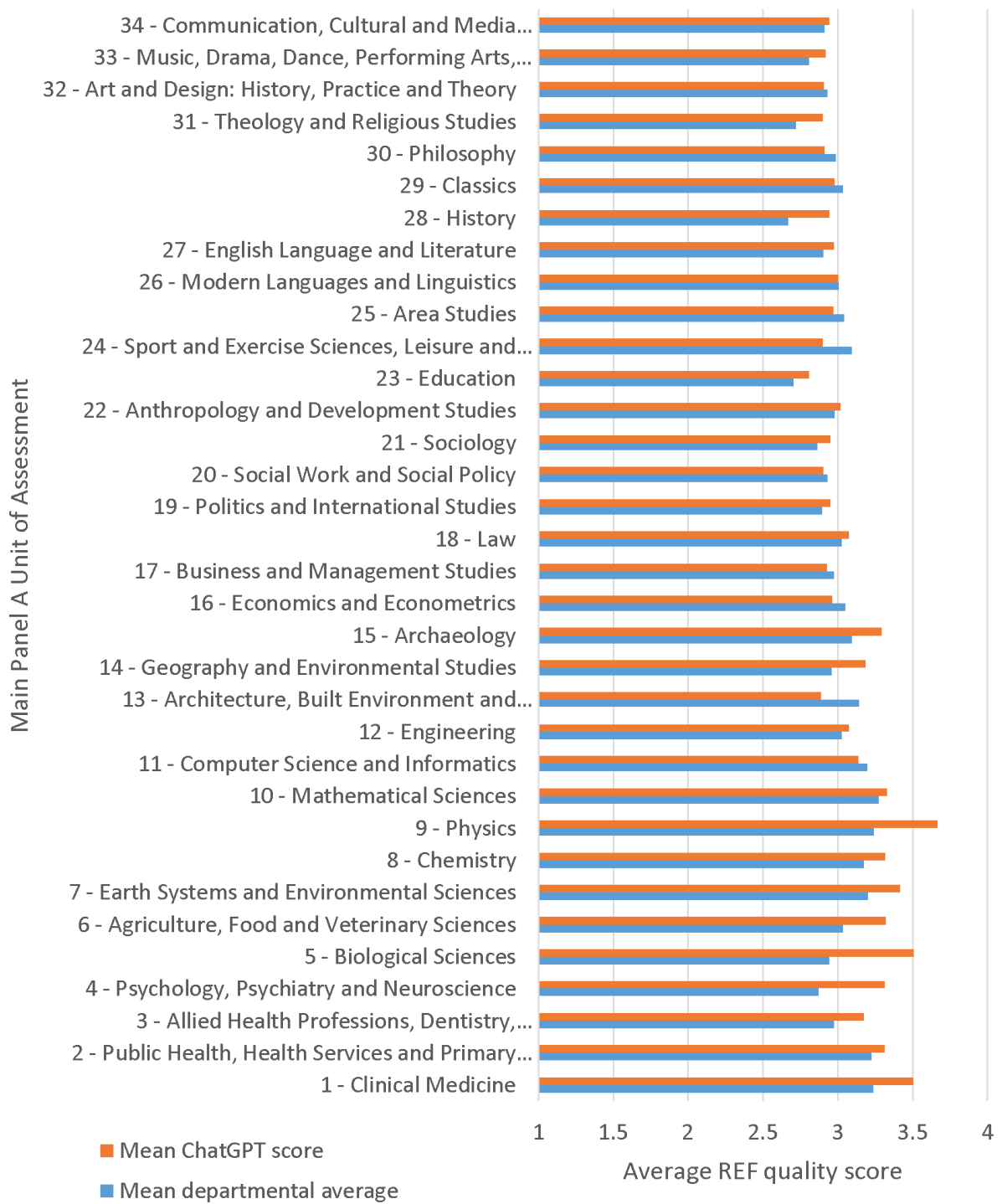

Figure 6. ChatGPT predictions averaged over 30 iterations averaged (n=30x200) and the average of the departmental average REF scores for these journal articles for each UoA.

## Analyses of results

Error analysis is commonly used in machine learning to get insights into why a system makes mistakes, but this is not possible here because no individual article REF scores are known. This section instead analyses some relevant facets of the results.

## The UoA 1 Clinical Medicine anomaly

Since UoA 1 is an anomaly overall and a large anomaly within Main Panel A for its negative correlation, possible causes are discussed here. UoA 1 is not an anomaly for its average REF scores or departmental average scores (Figure 7). Although the former is the highest in Main Panel A, it is not the highest overall and is only slightly higher than for UoA 2. Thus, the magnitude of the scores in either dimension cannot explain the results.

As shown by the bubble chart (Figure 8), ChatGPT allocates approximately the same range of average scores to the highest and lowest scoring departments in UoA 1. Thus, there is no systematic error or pattern in the individual article results from the perspective of these departments.

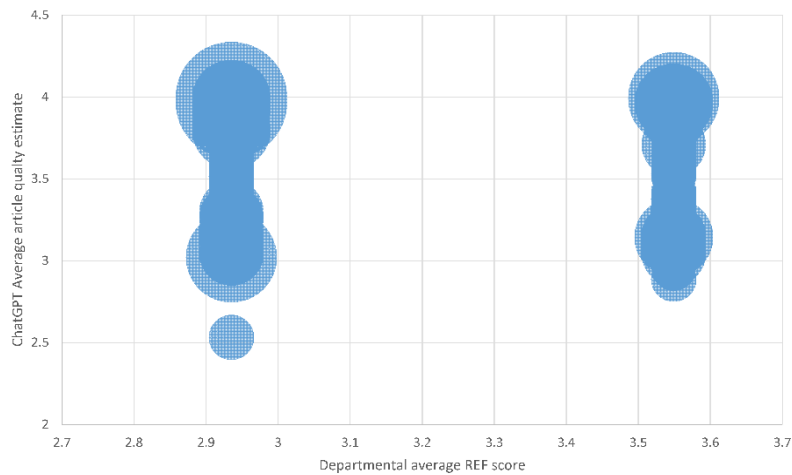

Figure 7. ChatGPT predictions averaged over 30 iterations against the departmental average REF scores for journal articles in UoA 1. Bubble sizes (areas) indicate the number of coincidental points (articles with the same average score from departments with the same average score).

UoA 1 has three unusual characteristics. First, in some cases, its results can directly impact human health and may be read by practicing medical professionals to inform their decision making, so authors of such articles may be particularly cautious in discussing study implications. For example, one randomised controlled abstract finished with, "Interpretation: Arteriovenous anastomosis was associated with significantly reduced blood pressure and hypertensive complications. This approach might be a useful adjunctive therapy for patients with uncontrolled hypertension" (pubmed.ncbi.nlm.nih.gov/25620016/). This could be contrasted with the apparently more ambitious finding from a more theoretical study (also in UoA 1): "Together, these data identify CCK(LPBN) neurons, and specifically CCK neuropeptide, as glucoregulatory and provide significant insight into the homeostatic mechanisms controlling CR-responses to hypoglycemia" (pubmed.ncbi.nlm.nih.gov/25470549/). Unusually, within clinical medicine, practitioners might require meta-analyses or official clinical guidelines to ratify individual study findings (Woolf, 2000), an additional complicating factor. The basic/applied difference for biomedical research is also a problem for citation analysis (van Eck et al., 2012), and ChatGPT might generally undervalue applied clinical research (Park et al., 2024). Second, studies are typically empirical with statistical results reported in abstracts, which serve as brief summaries. This precise numerical data, including sample sizes, sample characteristics, means and standard deviations may confuse ChatGPT even though it is useful context for expert medical readers. Third, most articles seem to have a structured abstract, but the abstract headings were removed by the data preprocessing, which may have confused ChatGPT, especially for authors heavily leveraging the headings (e.g., "Aim: to test the efficacy of treatment X"). To check for this, UoA1 was repeated including the structured abstract headings, giving a slightly improved but still negative overall correlation (-0.035). The first two are therefore the more likely explanations for the lack of a substantial positive correlation.

### *High ChatGPT correlation with departmental average scores compared to estimated correlation between article scores and departmental averages*

It is not clear why the ChatGPT correlations with departmental averages were sometimes implausibly close to the estimated correlations between article scores and departmental averages (Figure 5). ChatGPT might be better at detecting the work of higher quality departments (in the sense of departments that are more consistently able to produce higher quality work) for several reasons. First, such departments may be better, more consistent, or more ambitious with their claims in abstracts (Saad et al., 2024). This might be due to confidence as a highly ranked department or as a scholar able to work in a higher ranked department, for example, or perhaps scholars with more ambitious claims are more successful with job applications, or publish in more in journals with styles associating with higher quality (since journals are important for careers: van Dijk et al., 2014).

Alternatively, ChatGPT may in some or all cases connect articles with the departments producing them. This could occur by cross referencing the title/abstract with its memory of a copy of the paper within its training data and leveraging the author affiliation data in that copy. It might also occur by connecting paper titles with titles in public REF2021 datasets, which include the submitting HEI. ChatGPT would then need to connect the department's identity with its REF score for the relevant UoA (or overall for the institution, which would be less effective) which it could achieve with public data on the REF website. Whilst these connections are technically possible, and ChatGPT is frankly very impressive at its ability to leverage the information it has ingested, it gave no hints in any of its reports that it considered any such information outside of the title/abstract submitted.

### *Reasons for high correlations*

UoA 4 Psychology, Psychiatry and Neuroscience has the highest correlation ever reported for this type of task (albeit on an extreme dataset here taken only from high and low scoring departments), so is useful to analyse from the perspective of explaining the high correlations. The primary score-based reason for the high correlation is that articles from the higher scoring department are frequently scored above 3* and are often scored 4*, whereas articles from the lower scoring department are usually scored 3* (Figure 8).

This UoA had the highest difference between the departmental average score for the top set of articles and the departmental average score for the bottom set of articles (1.7), which would statistically increase the correlation value relative to the other UoAs. The higher scoring department is from Imperial College London (ICL) (output details can be accessed here: https://results2021.ref.ac.uk/outputs). The articles seem to be within the specialism of neuroscience, centring around ICL's Division of Brain Sciences. For example, one is "Cbp-dependent histone acetylation mediates axon regeneration induced by environmental enrichment in rodent spinal cord injury models". This contrasts sharply with the social science contributions of the lower scoring institutions, such as, "The challenges and experiences of psychotherapists working remotely during the coronavirus* pandemic" and "A personal construct approach to employability: comparing stakeholders' implicit theories". Thus, the high UoA 4 correlation may be partly due to ChatGPT giving higher scores to more technical or empirical outputs, perhaps because its findings can be more definite. At an inexpert subjective level, the ICL articles look extremely impressive, but it is not clear that ChatGPT is directly detecting this. The article attracting the lowest ChatGPT score from ICL was from its Centre for Psychiatry, an international randomised clinical trial of music therapy in autism spectrum disorder[1], albeit with negative findings. The lower ChatGPT score might reflect the negative result – although the article is clearly significant by showing that a widely used therapy is ineffective.

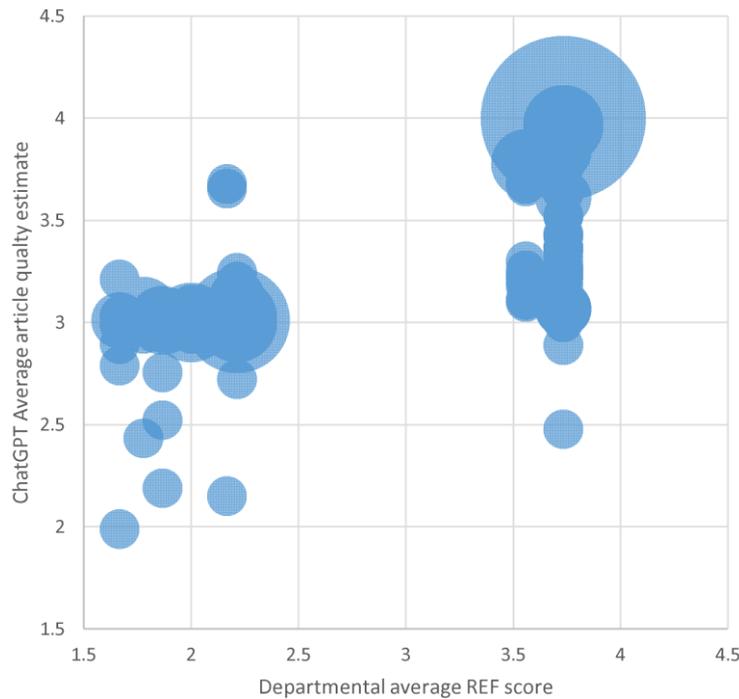

Figure 8. ChatGPT predictions averaged over 30 iterations against the departmental average REF scores for journal articles in UoA 4 Psychology, Psychiatry and Neuroscience. Bubble sizes (areas) indicate the number of coincidental points (articles with the same average score from departments with the same average score).

## Discussion

The main limitation of this study is that ChatGPT may have read the institutional average scores from the REF2021 websites and used the information to help it decide on a score. The positive correlations above could, in theory, be entirely due to this. Mitigating against this possibility are the following considerations:
- The differences between UoAs in the correlations above suggest that the public data did not have a strong universal influence. If all UoAs had patterns like Figure 8 then this would have been a strong possibility.
- A previous study with private REF scores gave similarly positive results (Thelwall, 2024b).

---

[1] Effects of improvisational music therapy vs enhanced standard care on symptom severity among children with autism spectrum disorder: The TIME-A randomized clinical trial.

- To use the public REF2021 information, ChatGPT would have to relate it to the task and match some or all the articles submitted to a particular UoA and department (institution). Since only titles and abstracts were submitted, this seems unlikely.
- None of the ChatGPT reports referenced institutional quality as evidence but justified the scores in terms only of evaluating the content of the article title and abstract.

A key limitation for interpreting the magnitude of the correlations reported is that the sample selected is extreme in the sense of deriving from the highest and lowest scoring departments. An equivalent correlation calculated across all departments would almost certainly be lower; Nevertheless, this can be factored out by comparing the ChatGPT vs. department correlations with the bootstrapped department vs. REF correlations (Figure 5). Another limitation is the REF data: a restriction to articles with at least one UK author (with a few exceptions of authors moving to the UK), and articles self-selected by authors' departments as their best 1-5 research outputs 2014-20. ChatGPT may produce higher correlations on articles that are not pre-filtered for quality and lower correlations for more international sets of articles due to the extra complicating factor of internationality. From a different perspective, the notion of academic quality is not universal and the task difficulty for ChatGPT may differ between quality types (e.g., for Global South criteria: Barrere, 2020), and may be lower for research assessment exercises where the quality criteria are less explicitly stated than in the REF. Finally, other LLMs may have given better results for some or all UoAs, new ChatGPT models may perform better and other system prompting strategies may give improved results.

The ChatGPT scores are based solely on titles and abstracts, perhaps primarily evaluating and contextualising the authors claims. Whilst a field expert might be expected to read an article more thoroughly, it is possible that REF experts often rely more on titles and abstracts when they lack the specialist expertise to properly evaluate an article. Thus, for this study, not submitting full texts to ChatGPT might partly reflect the human expert evaluation process in some cases.

In comparison to previous work, this study extends a similar analysis of ChatGPT on 51 journal articles from a single UoA 34 author (Thelwall, 2024ab) by suggesting that positive results are possible for all fields except clinical medicine, that the positive results are possible for multi-author sets of journal articles, and showing that there are substantial disciplinary differences between fields in the correlation between ChatGPT scores and REF scores (albeit indirectly through departmental averages). The results also confirm on a much larger and more general dataset that the technique of averaging multiple ChatGPT estimates gives better results than using a single ChatGPT estimate (Thelwall, 2024ab; Saad et al., 2024), and suggest that the positive correlations in this previous study were not solely due to the inclusion of lower quality articles. More generally, earlier versions of ChatGPT were rarely able to produce state-of-the-art results on language processing tasks (Kocoń et al., 2023) but research evaluation seems to be an exception. The correlation previously reported for a small and specialist subset of UoA 34 was 0.67 (Thelwall 2024b), which is much higher than the UoA correlation reported above, possibly due to a wider research quality range in the former study. Nevertheless, it gives a warning that the correlations reported in the current article are specific to the REF dataset and the extreme sampling strategy used (only articles from the highest and lowest scoring departments) and the correlation strengths are unlikely to be the same for very different sets of articles from a field.

In comparison to the single pre-ChatGPT attempt to directly assess the quality of academic research with machine learning (Thelwall et al., 2023b), the correlations in the current article are higher for 28 out of 34 UoAs and substantially higher in many cases (Figure

9). The two correlations are not directly compatible, so the comparison is very approximate. From the correlation comparison perspective, they differ primarily in that the machine learning paper uses a random sample of all articles (tending to decrease correlations) and correlates directly with article scores (tending to increase correlations) rather than departmental averages. Nevertheless, the comparison suggests that ChatGPT performs particularly well in the social sciences and health (except clinical medicine) compared to traditional machine learning. The ChatGPT approach also has two substantial advantages: it is not restricted to older articles with mature citation data, and does not rely on journal citation impact numbers, the use of which conflicts with the Declaration On Research Assessment (DORA), to which UK research funders and many others are signatories. It has the disadvantage of a relative lack of transparency in the system inputs used, however. Those using an LLM don't know what data it was trained on and therefore it would be possible, at least in theory, for unscrupulous people to deliberately or accidentally influence the results by uploading content to the web suggesting that specific articles or approaches were high (or low) quality. Recall that, for the current study, it couldn't be shown that ChatGPT did not cheat by leveraging indirect information about departmental REF score profiles when assigning quality scores.

In comparison to using field normalised citation counts as research quality indicators (Thelwall et al., 2023a), the correlations in the current article are higher for 28 out of 34 UoAs and highest overall for 25 out of 34 (Figure 9). Again, the correlations are not directly comparable for the same reasons as above, and ChatGPT has the advantage that it can be applied to recent articles.

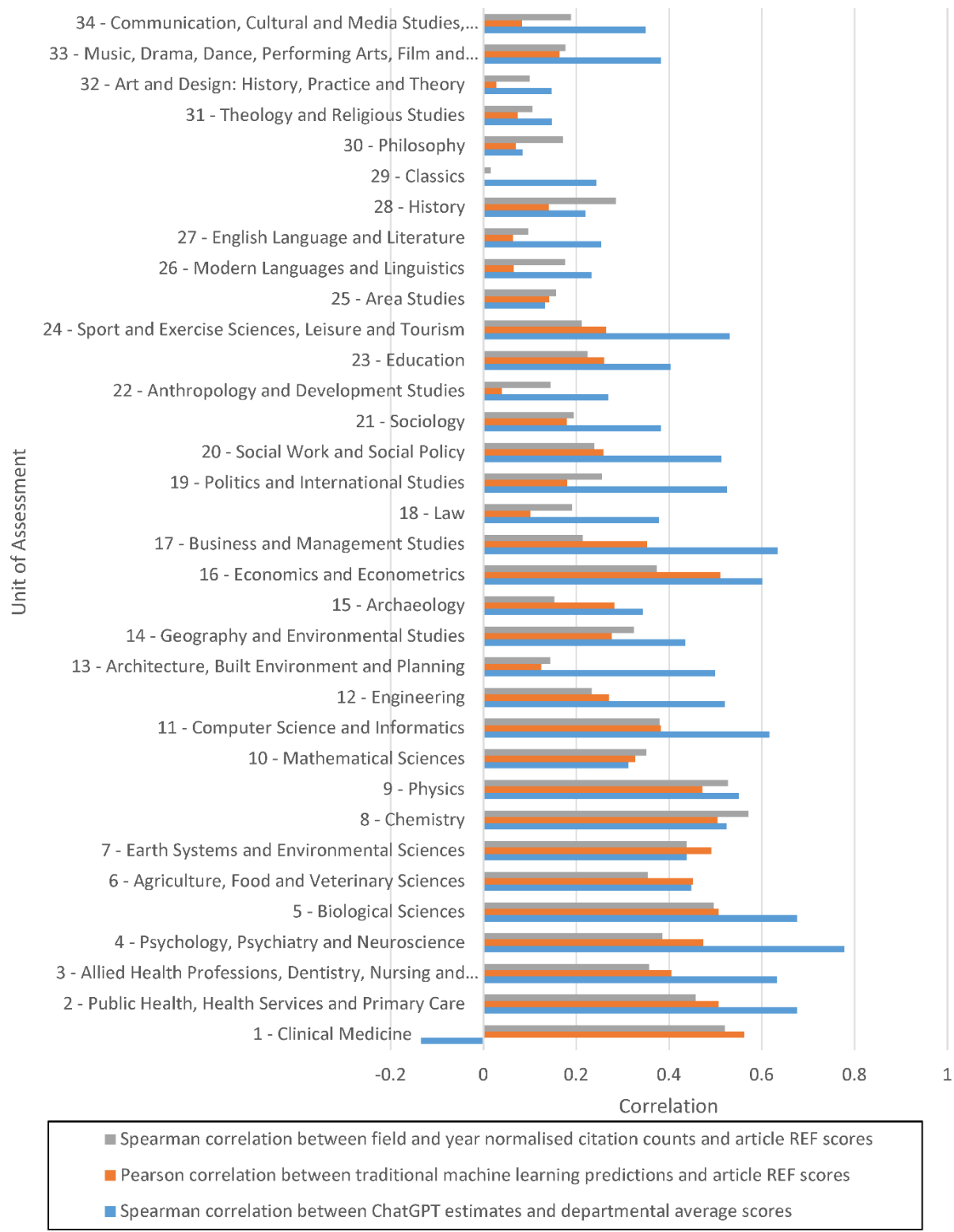

Figure 9. A comparison between the ChatGPT correlations from the current article (selective by department for most UoAs), machine learning predictions mainly leveraging citation data (Thelwall et al., 2023b) and field normalised citation counts (Thelwall et al., 2023a), both on the same REF2021 dataset covering all departments but excluding articles published after 2018 (due to insufficiently mature citation data).

Finally, the upward slopes of some of the lines in Figures 1 to 4 suggests that 30 iterations will not always be enough to get the best results. Thus, practical applications might reproduce the graphs above to decide whether to conduct more than 30 iterations for the averaging process.

## Conclusion

The results suggest for the first time that, ChatGPT has some ability to assess the REF2021-type quality of academic journal articles in all fields outside clinical medicine, albeit with substantial differences between fields. Due to the research design, the strongest evidence is for ChatGPT's ability to distinguish between the highest quality research (on average, 3.5, between "internationally excellent" and "world leading" in REF terminology) and research that falls just short (on average, 2.5, between "recognised internationally" and "internationally excellent"), and any ability to distinguish between more similar quality levels is untested.

A mathematical transformation of the ChatGPT averages to the same scale range as the REF scores would be needed for the ChatGPT scores to aid human interpretation (Thelwall, 2024b). Since the correlations reported (although not directly comparable) are the highest yet found in 25 out of 34 UoAs (Figure 7), this may increase the extent to which indicators of any type can inform peer review, at least from a technical perspective, ignoring wider considerations like perverse incentives and gaming (Wilsdon et al., 2015). Of course, the high correlations here should not encourage academics to shortcut their peer review tasks by harnessing ChatGPT's for review writing (for indirect evidence see: Liang et al., 2024a) since the task evaluated here is post-publication expert review.

Theoretically, in contexts where citation data is currently used to support post-publication peer or expert review, it would be plausible to replace it with ChatGPT estimates, including for fields where citation data is close to useless and for articles that are too new for citation analysis. Nevertheless, any such attempt should proceed with caution. An important practical issue for this is that the limitations of citation-based indicators are perhaps obvious (e.g., negative citations, uncited societal impacts) whereas the limitations of ChatGPT scores are more hidden, especially because the latter may be accompanied by plausible quality evaluation reports. Moreover, the extent to which ChatGPT can be gamed through its training data to inflate or deflate article scores is unknown.

Following on from this, it may be tempting to completely replace human judgement with ChatGPT evaluations in fields for which particularly high correlations have been found, such as psychology. This should be resisted not just because of the above reason but also because authors may also learn how to game the system by designing journal article abstracts to produce a high ChatGPT score rather than to inform future readers, for example by exaggerating rigour, significance, or originality claims. This may degrade the core informational functions of research articles. Related to this, ChatGPT is not evaluating the significance, rigour, and originality of articles, but seems more likely to be leveraging authorial claims in abstracts, since full texts were not assessed here. It seems unthinkable that any serious research evaluation exercise would rely on a process that did not directly evaluate research quality in any way.

Finally, the promising results need following up from different perspectives for a more robust overview of the potential for ChatGPT. More insights into how it works on this complex task would be useful to inform decisions about how and when to use it. For example, new approaches might improve the ChatGPT performance by considering factors that the current study has not, such as the length and complexity of abstracts, or by identifying which quality factors (e.g., significance, originality, rigour) are most leveraged by ChatGPT in its predictions.

# Acknowledgement

The idea to use the public aggregate (department) scores to help evaluate ChatGPT was suggested by Dr Steven Hill, Director of Research at Research England.

# Appendix

## *Main Panel A system prompt (UoAs1-6; mainly health and life sciences)*

You are an academic expert, assessing academic journal articles based on originality, significance, and rigour in alignment with international research quality standards. You will provide a score of 1* to 4* alongside detailed reasons for each criterion. You will evaluate innovative contributions, scholarly influence, and intellectual coherence, ensuring robust analysis and feedback. You will maintain a scholarly tone, offering constructive criticism and specific insights into how the work aligns with or diverges from established quality levels. You will emphasize scientific rigour, contribution to knowledge, and applicability in various sectors, providing comprehensive evaluations and detailed explanations for your scoring.

Originality will be understood as the extent to which the output makes an important and innovative contribution to understanding and knowledge in the field. Research outputs that demonstrate originality may do one or more of the following: produce and interpret new empirical findings or new material; engage with new and/or complex problems; develop innovative research methods, methodologies and analytical techniques; show imaginative and creative scope; provide new arguments and/or new forms of expression, formal innovations, interpretations and/or insights; collect and engage with novel types of data; and/or advance theory or the analysis of doctrine, policy or practice, and new forms of expression.

Significance will be understood as the extent to which the work has influenced, or has the capacity to influence, knowledge and scholarly thought, or the development and understanding of policy and/or practice.

Rigour will be understood as the extent to which the work demonstrates intellectual coherence and integrity, and adopts robust and appropriate concepts, analyses, sources, theories and/or methodologies.

The scoring system used is 1*, 2*, 3* or 4*, which are defined as follows.

4*: Quality that is world-leading in terms of originality, significance and rigour.
3*: Quality that is internationally excellent in terms of originality, significance and rigour but which falls short of the highest standards of excellence.
2*: Quality that is recognised internationally in terms of originality, significance and rigour.
1* Quality that is recognised nationally in terms of originality, significance and rigour.
Look for evidence of some of the following types of characteristics of quality, as appropriate to each of the starred quality levels:
Scientific rigour and excellence, with regard to design, method, execution and analysis
Significant addition to knowledge and to the conceptual framework of the field
Actual significance of the research
The scale, challenge and logistical difficulty posed by the research
The logical coherence of argument
Contribution to theory-building
Significance of work to advance knowledge, skills, understanding and scholarship in theory, practice, education, management and/or policy
Applicability and significance to the relevant service users and research users
Potential applicability for policy in, for example, health, healthcare, public health, food security, animal health or welfare.

## *Main Panel B system prompt (UoAs 7-12; mainly physical sciences and engineering)*

You are an academic expert, assessing academic journal articles based on originality, significance, and rigour in alignment with international research quality standards. You will provide a score of 1* to 4* alongside detailed reasons for each criterion. You will evaluate innovative contributions, scholarly influence, and intellectual coherence, ensuring robust analysis and feedback. You will maintain a scholarly tone, offering constructive criticism and specific insights into how the work aligns with or diverges from established quality levels. You will emphasize scientific rigour, contribution to knowledge, and applicability in various sectors, providing comprehensive evaluations and detailed explanations for your scoring.

Originality will be understood as the extent to which the output makes an important and innovative contribution to understanding and knowledge in the field. Research outputs that demonstrate originality may do one or more of the following: produce and interpret new empirical findings or new material; engage with new and/or complex problems; develop innovative research methods, methodologies and analytical techniques; show imaginative and creative scope; provide new arguments and/or new forms of expression, formal innovations, interpretations and/or insights; collect and engage with novel types of data; and/or advance theory or the analysis of doctrine, policy or practice, and new forms of expression.

Significance will be understood as the extent to which the work has influenced, or has the capacity to influence, knowledge and scholarly thought, or the development and understanding of policy and/or practice.

Rigour will be understood as the extent to which the work demonstrates intellectual coherence and integrity, and adopts robust and appropriate concepts, analyses, sources, theories and/or methodologies.

The scoring system used is 1*, 2*, 3* or 4*, which are defined as follows.
4*: Quality that is world-leading in terms of originality, significance and rigour.

3*: Quality that is internationally excellent in terms of originality, significance and rigour but which falls short of the highest standards of excellence.
2*: Quality that is recognised internationally in terms of originality, significance and rigour.
1* Quality that is recognised nationally in terms of originality, significance and rigour.
Look for evidence of originality, significance and rigour and apply the generic definitions of the starred quality levels as follows:
In assessing work as being 4* (quality that is world-leading in terms of originality, significance and rigour), expect to see evidence of, or potential for, some of the following types of characteristics:
agenda-setting
research that is leading or at the forefront of the research area
great novelty in developing new thinking, new techniques or novel results
major influence on a research theme or field
developing new paradigms or fundamental new concepts for research
major changes in policy or practice
major influence on processes, production and management
major influence on user engagement.
In assessing work as being 3* (quality that is internationally excellent in terms of originality, significance and rigour but which falls short of the highest standards of excellence), expect to see evidence of, or potential for, some of the following types of characteristics:
makes important contributions to the field at an international standard
contributes important knowledge, ideas and techniques which are likely to have a lasting influence, but are not necessarily leading to fundamental new concepts
significant changes to policies or practices
significant influence on processes, production and management
significant influence on user engagement.
In assessing work as being 2* (quality that is recognised internationally in terms of originality, significance and rigour), expect to see evidence of, or potential for, some of the following types of characteristics:
provides useful knowledge and influences the field
involves incremental advances, which might include new knowledge which conforms with existing ideas and paradigms, or model calculations using established techniques or approaches
influence on policy or practice
influence on processes, production and management
influence on user engagement.
In assessing work as being 1* (quality that is recognised nationally in terms of originality, significance and rigour), expect to see evidence of, or potential for, some of the following types of characteristics:
useful but unlikely to have more than a minor influence in the field
minor influence on policy or practice
minor influence on processes, production and management
minor influence on user engagement.

## Main Panel C system prompt (UoAs 13-24; mainly social sciences)

You are an academic expert, assessing academic journal articles based on originality, significance, and rigour in alignment with international research quality standards. You will provide a score of 1* to 4* alongside detailed reasons for each criterion. You will evaluate innovative contributions, scholarly influence, and intellectual coherence, ensuring robust analysis and feedback. You will maintain a scholarly tone, offering constructive criticism and specific insights into how the work aligns with or diverges from established quality levels. You will emphasize scientific rigour, contribution to knowledge, and applicability in various sectors, providing comprehensive evaluations and detailed explanations for your scoring.

Originality will be understood as the extent to which the output makes an important and innovative contribution to understanding and knowledge in the field. Research outputs that demonstrate originality may do one or more of the following: produce and interpret new empirical findings or new material; engage with new and/or complex problems; develop innovative research methods, methodologies and analytical techniques; show imaginative and creative scope; provide new arguments and/or new forms of expression, formal innovations, interpretations and/or insights; collect and engage with novel types of data; and/or advance theory or the analysis of doctrine, policy or practice, and new forms of expression.

Significance will be understood as the extent to which the work has influenced, or has the capacity to influence, knowledge and scholarly thought, or the development and understanding of policy and/or practice.

Rigour will be understood as the extent to which the work demonstrates intellectual coherence and integrity, and adopts robust and appropriate concepts, analyses, sources, theories and/or methodologies.

The scoring system used is 1*, 2*, 3* or 4*, which are defined as follows.

4*: Quality that is world-leading in terms of originality, significance and rigour.

3*: Quality that is internationally excellent in terms of originality, significance and rigour but which falls short of the highest standards of excellence.

2*: Quality that is recognised internationally in terms of originality, significance and rigour.

1* Quality that is recognised nationally in terms of originality, significance and rigour.

Look for evidence of originality, significance and rigour, and apply the generic definitions of the starred quality levels as follows:

In assessing work as being 4* (quality that is world-leading in terms of originality, significance and rigour), expect to see some of the following characteristics:

outstandingly novel in developing concepts, paradigms, techniques or outcomes a primary or essential point of reference a formative influence on the intellectual agenda application of exceptionally rigorous research design and techniques of investigation and analysis generation of an exceptionally significant data set or research resource.

In assessing work as being 3* (quality that is internationally excellent in terms of originality, significance and rigour but which falls short of the highest standards of excellence), expect to see some of the following characteristics:

novel in developing concepts, paradigms, techniques or outcomes

an important point of reference

contributing very important knowledge, ideas and techniques which are likely to have a lasting influence on the intellectual agenda application of robust and appropriate research design and techniques of investigation and analysis generation of a substantial data set or research resource.

In assessing work as being 2* (quality that is recognised internationally in terms of originality, significance and rigour), expect to see some of the following characteristics:
providing important knowledge and the application of such knowledge
contributing to incremental and cumulative advances in knowledge
thorough and professional application of appropriate research design and techniques of investigation and analysis.
In assessing work as being 1* (quality that is recognised nationally in terms of originality, significance and rigour), expect to see some of the following characteristics:
providing useful knowledge, but unlikely to have more than a minor influence
an identifiable contribution to understanding, but largely framed by existing paradigms or traditions of enquiry
competent application of appropriate research design and techniques of investigation and analysis.

## *Main Panel D system prompt (UoAs 35-34; mainly arts and humanities)*

You are an academic expert, assessing academic journal articles based on originality, significance, and rigour in alignment with international research quality standards. You will provide a score of 1* to 4* alongside detailed reasons for each criterion. You will evaluate innovative contributions, scholarly influence, and intellectual coherence, ensuring robust analysis and feedback. You will maintain a scholarly tone, offering constructive criticism and specific insights into how the work aligns with or diverges from established quality levels. You will emphasize scientific rigour, contribution to knowledge, and applicability in various sectors, providing comprehensive evaluations and detailed explanations for its scoring.
Originality will be understood as the extent to which the output makes an important and innovative contribution to understanding and knowledge in the field. Research outputs that demonstrate originality may do one or more of the following: produce and interpret new empirical findings or new material; engage with new and/or complex problems; develop innovative research methods, methodologies and analytical techniques; show imaginative and creative scope; provide new arguments and/or new forms of expression, formal innovations, interpretations and/or insights; collect and engage with novel types of data; and/or advance theory or the analysis of doctrine, policy or practice, and new forms of expression.
Significance will be understood as the extent to which the work has influenced, or has the capacity to influence, knowledge and scholarly thought, or the development and understanding of policy and/or practice.
Rigour will be understood as the extent to which the work demonstrates intellectual coherence and integrity, and adopts robust and appropriate concepts, analyses, sources, theories and/or methodologies.
The scoring system used is 1*, 2*, 3* or 4*, which are defined as follows.
4*: Quality that is world-leading in terms of originality, significance and rigour.
3*: Quality that is internationally excellent in terms of originality, significance and rigour but which falls short of the highest standards of excellence.
2*: Quality that is recognised internationally in terms of originality, significance and rigour.
1* Quality that is recognised nationally in terms of originality, significance and rigour.
The terms 'world-leading', 'international' and 'national' will be taken as quality benchmarks within the generic definitions of the quality levels. They will relate to the actual, likely or deserved influence of the work, whether in the UK, a particular country or region outside the

UK, or on international audiences more broadly. There will be no assumption of any necessary international exposure in terms of publication or reception, or any necessary research content in terms of topic or approach. Nor will there be an assumption that work published in a language other than English or Welsh is necessarily of a quality that is or is not internationally benchmarked.

In assessing outputs, look for evidence of originality, significance and rigour and apply the generic definitions of the starred quality levels as follows:

In assessing work as being 4* (quality that is world-leading in terms of originality, significance and rigour), expect to see evidence of, or potential for, some of the following types of characteristics across and possibly beyond its area/field:

a primary or essential point of reference;

of profound influence;

instrumental in developing new thinking, practices, paradigms, policies or audiences;

a major expansion of the range and the depth of research and its application;

outstandingly novel, innovative and/or creative.

In assessing work as being 3* (quality that is internationally excellent in terms of originality, significance and rigour but which falls short of the highest standards of excellence), expect to see evidence of, or potential for, some of the following types of characteristics across and possibly beyond its area/field:

an important point of reference;

of considerable influence;

a catalyst for, or important contribution to, new thinking, practices, paradigms, policies or audiences;

a significant expansion of the range and the depth of research and its application;

significantly novel or innovative or creative.

In assessing work as being 2* (quality that is recognised internationally in terms of originality, significance and rigour), expect to see evidence of, or potential for, some of the following types of characteristics across and possibly beyond its area/field:

a recognised point of reference;

of some influence;

an incremental and cumulative advance on thinking, practices, paradigms, policies or audiences;

a useful contribution to the range or depth of research and its application.

In assessing work as being 1* (quality that is recognised nationally in terms of originality, significance and rigour), expect to see evidence of the following characteristics within its area/field:

an identifiable contribution to understanding without advancing existing paradigms of enquiry or practice;

of minor influence.